\documentclass{article}

\usepackage[preprint]{corl_2022} % Uncomment for pre-prints (e.g., arxiv); This is like ``final'', but will remove the CORL footnote.
\usepackage{times}  % DO NOT CHANGE THIS
\usepackage{helvet}  % DO NOT CHANGE THIS
\usepackage{courier}  % DO NOT CHANGE THIS
\usepackage{graphicx} % DO NOT CHANGE THIS
\usepackage{natbib}  % DO NOT CHANGE THIS AND DO NOT ADD ANY OPTIONS TO IT
\usepackage{caption} % DO NOT CHANGE THIS AND DO NOT ADD ANY OPTIONS TO IT
\usepackage{algorithm}
\usepackage[noend]{algpseudocode}

\usepackage{adjustbox}
\usepackage{wrapfig}
\usepackage{amsmath}
\usepackage{amsthm}
\usepackage{mathtools}
\usepackage{mathptmx}
\usepackage{amssymb}
\usepackage{dsfont}
\usepackage{enumitem}

\usepackage{newfloat}
\usepackage{listings}
\DeclareCaptionStyle{ruled}{labelfont=normalfont,labelsep=colon,strut=off} % DO NOT CHANGE THIS
\floatstyle{ruled}
\newfloat{listing}{tb}{lst}{}
\floatname{listing}{Listing}
\DeclareMathOperator*{\sbjto}{s.\ t.\ }

\DeclareMathOperator{\rank}{rank}
\DeclareMathOperator{\sat}{sat}
\DeclareMathOperator{\trace}{tr}
\DeclareMathOperator{\proj}{Proj}

\renewcommand{\leq}{\leqslant}

\renewcommand{\geq}{\geqslant}

\newcommand{\R}{\mathds{R}}
\newcommand{\Nz}{\mathds{N}_0}
\newcommand{\N}{\mathds{Z}_{+}}

\newcommand{\bmat}[1]{\begin{bmatrix}#1\end{bmatrix}}
\newcommand{\abs}[1]{\left|#1\right|}
\newcommand{\norm}[1]{\left\|#1\right\|}
\newcommand{\secref}[1]{\S \ref{#1}}

\newcommand{\transp}{^\top}

\newcommand{\zeros}{\mathbf{0}}

\newcommand{\st}{x}

\newcommand{\control}{u}
\newcommand{\controlset}{\mathds{U}}
\newcommand{\wnoise}{w}

\newcommand{\costps}{c_{\mathrm{s}}}
\newcommand{\costfinal}{c_{\mathrm{f}}}

\newcommand{\authority}{u_{\max}}

\newcommand{\Let}{\coloneqq}
\newcommand{\teL}{\eqqcolon}

\newtheorem{assumption}{Assumption}
\newtheorem{remark}{Remark}
\newtheorem{theorem}{Theorem}
\newtheorem{lemma}{Lemma}
\newtheorem{proposition}{Proposition}

\newtheorem{definition}{Definition}

\allowdisplaybreaks

\title{Deep Model Predictive Control}

\author{
  Prabhat K.~Mishra\\
  UIUC, USA \\
  \texttt{pmishra@illinois.edu} \\
  \And
  Mateus V.~Gasparino\\
  UIUC, USA \\
  \texttt{mvalve2@illinois.edu} \\
  \And
  Andres E.~B.~Velasquez\\
  UIUC, USA \\
   \texttt{andresbaquerov@gmail.com} \\
  \And
  Girish Chowdhary\\
  UIUC, USA \\
  \texttt{girishc@illinois.edu}
}

\begin{document}
\maketitle

%===============================================================================

\begin{abstract}
    This paper presents a deep learning based model predictive control algorithm for control affine nonlinear discrete time systems with matched and bounded state-dependent uncertainties of unknown structure. Since the structure of uncertainties is not known, a deep neural network (DNN) is employed to approximate the disturbances. In order to avoid any unwanted behavior during the learning phase, a tube based model predictive controller is employed, which ensures satisfaction of constraints and input-to-state stability of the closed-loop states. 
\end{abstract}


%===============================================================================

\section{Introduction}
	Modeling errors and environmental uncertainties are unavoidable in practice. Therefore, purely model based controllers tend to exhibit unexpected or unwanted behaviors in the real-world. One key solution to this problem is to employ learning-based methods that utilize powerful learning elements such as \emph{deep neural networks} (DNN). Such methods attempt to learn a good model of underlying nonlinear dynamics while the system is in operation in a manner that does not compromise safety and performance. We refer readers to \cite{lecun2015deep, Silver_Nature15, Levine_JMLR16, bojarski2016, survey_LMPC, robust_action_governor, safe_rl_Krause, robust_regression, safe_exploration, safe_exploration_Krause, safe_RL_Tomlin} and references therein. 
    
    \par To address the above challenge, the available domain knowledge in terms of approximate model is utilized in  \cite{Tomlin_IROS11, NN_collision_avoidance}, along with the learning elements. We refer readers to an excellent survey on safe reinforcement learning \cite{safeRL_survey} and references therein. One key approach for safe learning is to augment the learning based controller with \emph{model predictive control} (MPC) and related methods to guarantee safety through constraint satisfaction and improve the performance over time \cite{lbmpc_linear, LMPC_Borrelli, predictive_safety_filter, SLS_safe, learning_MPC_Mesbah, deepreach, Deep_MPC_Gopaluni, Deep_MPC_Lucia_18, rapid_MPC_Borreli, minimax_Deep_MPC, Allgower_MPC_approximate, NNMPC_Findeisen, Data-driven_Muller}. 
    The proper pairing of learning and MPC can bring useful features of both methods while compensating their drawbacks.

    \par Our main goal in this article is to address these gaps by creating a learning based MPC architectures with performance and safety guarantees. When uncertainties are structured, they can be simply represented in terms of (possibly) high dimensional feature basis functions and the learning mechanism acts on the disturbances \cite{MRAC_Nguyen, HHN_08, Chowdhary_GP_14, GPMRAC_18, DMRAC, Joshi_Virdi}. These disturbance rejecting actions taken by the learning mechanism are experienced by the MPC controller as additional disturbances. If the learning mechanism eventually rejects the disturbance then MPC can ensure asymptotic convergence of closed-loop states while satisfying the underlying constraints \cite{MWGC}. In this article, we extend the results of \cite{MWGC} for unstructured uncertainties.

We present a problem setup in \secref{s:setup}. The formulation of Deep MPC controller is given in \secref{s:MPC}. We validate our theoretical results with the help of a numerical experiments in \secref{s:experiment} and conclude in \secref{s:epilogue}. The real time implementable training mechanism of DNN, stability of the overall algorithm and proofs are given in the appendix. 	 

We let $\R, \Nz, \Nz$ denote the set of real numbers, non-negative integers and positive integers, respectively. For a given vector $v$ and positive (semi)-definite matrix $M \succeq \zeros$, $\norm{v}_M^2$ is used to denote $v \transp M v$. For a given matrix $A$, the trace, the largest eigenvalue, pseudo-inverse and Frobenius norm are denoted by $\trace(A)$, $\lambda_{\max}(A)$, $A^{\dagger}$ and $\norm{A}_F$, respectively. By notation $\norm{A}$ and $\norm{A}_{\infty}$, we mean the standard $2-$norm and $\infty-$norm, respectively, when $A$ is a vector, and induced $2-$norm and $\infty-$norm, respectively, when $A$ is a matrix. 
A vector or a matrix with all entries $0$ is represented by $\zeros$ and $I$ is an identity matrix of appropriate dimensions. We let $M^{(i)}$ denote the $i^{\text{th}}$ column of a given matrix $M$.

\section{Problem setup}\label{s:setup}

Let us consider a discrete time dynamical system 
\begin{equation}\label{e:system}
\st_{t+1} = f(\st_t) + g(\st_t)\left( \control_t + h(\st_t) \right), \text{ where }
\end{equation}
\begin{enumerate}[leftmargin = *, nosep, label=(1-\alph*), widest = b]
	\item \label{e:constraints} $\st_t \in \mathcal{X} \subset \R^d$, $ \control_t \in \controlset \Let \{v \in \R^m \mid \norm{v}_{\infty} \leq \authority \}$, $\mathcal{X}\subset \R^d$ is a compact set,
%	\item  
	\item system function $f:\R^d \rightarrow \R^d $, control influence function $g:\R^d \rightarrow \R^{d \times m}$ are given Lipschitz continuous functions and represent domain knowledge or prior knowledge of the system dynamics.
	\item \label{as: bounds_hg} $h(\st_t)$ is the state dependent matched uncertainty at time $t$ such that $g(\st_t)h(\st_t) \in \mathds{W} \Let \{v \in \R^d \mid \norm{v} \leq w_{\max} \}$, $h$ is continuous, $\norm{g(\st_t)} \leq \delta_g$ for some $\delta_g> 0$ and $\rank(g(\st_t)) = m$ for every $\st_t \in \R^d$. 
\end{enumerate}  
The term $f(\st_t) + g(\st_t) \control_t$ in the right hand side of \eqref{e:system} represents the prior knowledge of the dynamics and the remaining term $g(\st_t)h(\st_t)$ represents the unknown part of the dynamics or uncertainties. We refer readers to \cite{wing-rock-theis, NASA_2009, Lewis_NN_99, inverted_pendulum_friction, neuroadaptive_Yucelen, Pavon_unmatched} for a few related problem formulations.

\section{Deep Model predictive controller}\label{s:MPC}
Our proposed solution is based on constraint satisfaction and cost minimization capabilities of MPC, and universal approximation property of neural networks. We break the applied control $\control_t$ such that
\begin{equation}\label{e:total_conrol} 
\control_t = \control_t^a + \control_t^m, 
\end{equation}
where $\control_t^a$ is the output of DNN and $\control_t^m$ is the MPC components, at time $t$. The relevant details about DNN are given in the Appendix \secref{s:adaptive}. The MPC controller employs only the nominal dynamics of \eqref{e:system}, which is given below for easy reference
\begin{equation}\label{e:nominal}
\st_{t+1} = f(\st_t) + g(\st_t) \control_t^m  \Let \bar{f}(\st_t, \control_t^m).
\end{equation}
Therefore, the dynamics \eqref{e:system} can be written as
\begin{equation}\label{e:actual_agent}
\st_{t+1} = \bar{f}(\st_t, \control_t^m) + g(\st_t) \left( \control_t^a + h(\st_t)\right).
\end{equation}

\par Notice that in \eqref{e:actual_agent}, the term $g(\st_t) \left( \control_t^a + h(\st_t)\right)$ is independent of the MPC control component $\control_t^m$. Therefore, MPC experiences it as a disturbance. In a broader sense, the MPC component $\control_t^m$ is responsible for input-to-state stability (ISS) of closed-loop states in the presence of bounded disturbances, and the DNN component $\control_t^a$ acts on $h(\st_t)$. In particular, the job of $\control_t^a$ is to approximate $-h(\st_t)$ and keep the approximation error uniformly bounded with a known bound so that MPC can always experience a bounded disturbance. 

\par Deep MPC is developed on celebrated tube based MPC \cite{Mayne_tube_NLMPC} with some differences, which occur due to the inclusion of the DNN component $u_t^a$. Tube based MPC ensures that the closed-loop states stay within a tube around a reference trajectory. The trackable reference trajectory is obtained by solving a reference governor problem \emph{offline} under the tightened constraints for regulation problems. Once a trackable reference trajectory is obtained by spending only a part of the available control authority, a reference tracking problem without state constraints is solved \emph{online} that utilizes full control authority. 
\par Constraint tightening in the reference governor allows satisfaction of the actual constraints by the actual states and actual actions. Knowledge of the exact bound on disturbance, therefore, is needed to tighten the constraints. Although the disturbance in dynamical system \eqref{e:system} at time $t$ is $g(x_t)h(x_t)$, the disturbance experienced by MPC is $g(\st_t)(\control_t^a + h(x_t))$, which can be proved uniformly bounded by carefully designed DNN and its training mechanism. More details about getting the bounds $\norm{u_t^a} \leq \authority^a$ and $\norm{g(x_t)(u_t^a+h(x_t))}\leq w_{\max}^\prime$ are given in the Appendix \secref{s:outer layer training}. Therefore, we re-define the disturbance set and control set as follows:
\[ \mathds{W}^{\prime} \Let \{v \in \R^d \mid \norm{v} \leq w_{\max}^{\prime} \}, \quad \controlset^{\prime} \Let \{v \in \R^m \mid \norm{v}_{\infty} \leq \authority - \authority^a \} .\]

These modifications in tube-based MPC are already pointed out in \cite{MWGC, NNMPC_Findeisen, Bhasin_AMPC_19}.
For some optimization horizon $N \in \N$, an offline reference governor is utilized to generate a reference trajectory 
\begin{equation}\label{e:reference_signal}
\begin{aligned}
(\st_t^r)_{t \in \Nz} &\Let \{ (\st_t^r)_{t = 0}^{N-1}, 0, \ldots \}, \quad  
(\control_t^r)_{t \in \Nz} \Let \{ (\control_t^r)_{t = 0}^{N-1}, 0, \ldots \} .
\end{aligned}
\end{equation}
In particular, the reference trajectory \eqref{e:reference_signal} is obtained by solving the following optimal control problem with penalty matrices $Q,R \succ 0$ and tightened sets $\mathcal{X}_r, \controlset_r$:
\begin{equation}\label{e:reference_governor}
    \begin{aligned}
    \min_{(\control_i^r)_{i=0}^{N-1}} & \quad \sum_{i=0}^{N-1} \norm{\st_i^r}^2_Q + \norm{\control_i^r}_R^2  \\
    \sbjto & \quad \st_0^r = \st_0, \st_N^r = \zeros, \\
    & \quad \st_{i+1}^r = \bar{f}(\st_i^r, \control_i^r), \st_i^r \in \mathcal{X}_r \subset \mathcal{X},  \\
    & \quad \control_i^r \in \controlset_r \subset \controlset^{\prime}; i = 0, \ldots, N-1,  
    \end{aligned}
\end{equation}
where $\bar{f}$ is defined in \eqref{e:nominal}.
The tightened constraint sets $\mathcal{X}_r$ and $\controlset_r$ can be obtained by following the approach of \cite[\S 7]{Mayne_tube_NLMPC}. In order to design the online reference tracking MPC, we first choose an optimization horizon $N \in \N$ and positive definite matrices $Q, R \succ \zeros$, which can be different from those chosen for the reference governor. Let 
\[ \costps(\st_{t+i \mid t}, \control_{t+i \mid t}) \Let \norm{\st_{t+i \mid t} - \st_{t+i}^r}_Q^2 + \norm{\control_{t+i \mid t} - \control_{t+i}^r}_R^2\]
be the cost per stage at time $t+i$ predicted at time $t$ and let $\costfinal(x) \Let x\transp Q_f x$ be the terminal cost with $Q_f \succ 0$. The terminal cost $\costfinal$ is treated as a local control Lyapunov function within a terminal set 
\begin{equation}\label{e:terminal_set}
\mathcal{X}_f \Let \{x \in \R^d \mid \costfinal(x) \leq \alpha; \; \alpha > 0 \}
\end{equation}
as in \cite{Mayne_tube_NLMPC} by making the following assumption:
\begin{assumption}\label{as:stability}
	\rm{
		There exists a control $\control^{\prime} \in \controlset^{\prime} $ such that the following holds
		\begin{equation}
		\costfinal \left( \bar{f}(x, \control^{\prime}) \right) -\costfinal(x)  \leq -\costps(x, \control^{\prime}) \text{ for every } x \in \mathcal{X}_f.
		\end{equation}		
	}
\end{assumption}
The above assumption is standard in the literature. Refer to \cite[\S 4]{Mayne_tube_NLMPC} for more details with a minor modification, which we made here for simplicity.
Let us define 
\begin{equation}\label{e:cost_function}
V_m(\st_{t\mid t}, (\control_{t+i\mid t})_{i=0}^{N-1}) \Let \costfinal(\st_{t+N \mid t}) + \sum_{i=0}^{N-1}\costps(\st_{t+i \mid t}, \control_{t+i \mid t}) .
\end{equation}  
The online reference tracking MPC minimizes \eqref{e:cost_function} at each time instant $t$ under the following constraints:
\begin{align}
&\st_{t \mid t} = \st_t \label{e:constraint_initial}\\
& \control_{t \mid t} + \control_t^a \in \controlset \label{e:constraint_first_control}\\
& \st_{t+ i +1 \mid t} = \bar{f}(\st_{t+i \mid t}, \control_{t+i\mid t}) \text{ for } i = 0, \cdots, N-1   \label{e:constraint_dynamics}\\
& \control_{t + i\mid t} \in \controlset^\prime \text{ for } i = 1, \cdots, N-1 . \label{e:constraint_remaining_control}
\end{align}
Notice that the constraint \eqref{e:constraint_first_control} is different from the constraints present in the tube-based MPC formulation \cite{Mayne_tube_NLMPC}. We define the underlying optimal control problem as follows:
\begin{equation}\label{e:MPC}
V_m(\st_{t}) \Let \min_{(\control_{t+i \mid t})_{i=0}^{N-1}} \left\{ \eqref{e:cost_function} \mid \eqref{e:constraint_initial}, \eqref{e:constraint_first_control}, \eqref{e:constraint_dynamics}, \eqref{e:constraint_remaining_control} \right\}.
\end{equation}
Let the optimizer of the above problem be $(\control_{t+i\mid t}^\ast)_{i=0}^{N-1}$. Then the optimal cost will be $ V_m(\st_t) \Let V_m(\st_t, (\control_{t+i\mid t}^\ast )_{i=0}^{N-1} )$. 
The first control $\control_{t\mid t}^\ast$ is called the MPC component $\control_t^m$ and is applied along with $\control_t^a$ to the system at time $t$.
\section{Numerical experiment}\label{s:experiment}
\par We consider Wing-rock dynamics to corroborate our result. 
Letting $\delta_t$ denote the roll angle in radian, and $p_t$ denote the roll rate in radian per second, the state of the wing-rock dynamics model is $x_t \Let \bmat{\delta_t & p_t}\transp$ at time $t$. We consider the following discrete time dynamics: 
\begin{equation}\label{eq:disc-sys}
x_{t+1} = Ax_t + B\Big(u_t + h(x_t)\Big),
\end{equation}
where $A=\begin{bmatrix}
1 & 0.05 \\ 0 & 1
\end{bmatrix}$, $B=\begin{bmatrix}
0 \\ 0.05
\end{bmatrix}$,
and $h(\cdot)$ is bounded uncertainty. In order to generate $h$ for the purpose of simulation, we use $ h(x_t) = V_t\transp \varsigma(x_t) + \omega_t$, with
$V_t = v_t  V_0$, where
\begin{align*}
&v_t =  \begin{cases}
4 \text{ for } t \in [50\ell, 50\ell +49] \quad \ell = 0,2,4, \ldots \\
0  \text{ otherwise}.
\end{cases} \\
&V_0\transp = \bmat{0.8 & 0.2314 & 0.6918 & -0.6245 & 0.0095 & 0.0214},
\end{align*} 
and  $\omega_t \in [-\bar{\omega}, \bar{\omega}]$ is a truncated normal random variable with $\bar{\omega} = 0.1523$. 
The function $\varsigma(\cdot)$ is saturated by a standard saturation function as $\varsigma(x) = \sat(\varsigma^{\prime}(x))$, where $\varsigma^{\prime}(x) = \begin{bmatrix}
1 & \delta & p & \abs{\delta} p & \abs{p} p &\delta^3
\end{bmatrix}\transp $ and $\sat(\cdot)$ is a standard saturation function with the threshold $\frac{\bar{\omega}}{5}$. 
The controller is not aware of $\varsigma(\cdot)$ and $\omega$. The admissible state and control sets are given below:
\[ \mathcal{X}  = \left[-\frac{\pi}{6},\frac{\pi}{6} \right] \times \left[ -\frac{\pi}{3},\frac{\pi}{3} \right], \text{ and }
\controlset  = \left[-\frac{\pi}{4},\frac{\pi}{4} \right]. \]
\begin{wrapfigure}{l}{0.5\columnwidth}
	\centering
	\begin{adjustbox}{width = 0.5\columnwidth}
		\includegraphics{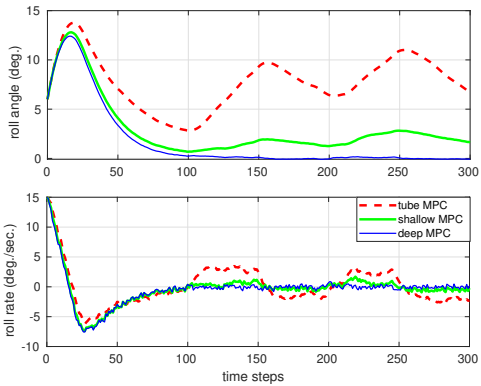}
	\end{adjustbox}
	\caption{deep MPC outperforms shallow MPC }
	\label{fig:experiment2}
	\vspace{-0.5cm}
\end{wrapfigure}
\par Our control objective is to steer the states of the system from $\st_0 = \bmat{\pi/30 & \pi/12}\transp$ to the origin.  We compare our proposed approach with two controllers, namely tube MPC \cite{Mayne_tube_NLMPC} and shallow MPC. In order to design shallow MPC, we follow our approach but we consider only a single layer neural network with $3$ neurons. 
To design the deep MPC, we use a four layer network with sizes $[2, 5, 5, 3]$ respectively, where the first hidden layer has 5 neurons and the outermost layer has $3$ neurons. The weights of the output layer are updated with our adaptive weight update law (\secref{s:outer layer training}), while the remaining three hidden layers are trained on a secondary machine (\secref{s:inner layer training}) using SGD with momentum constant 0.9 and learning rate 0.01. We use nonlinear activation functions after each of the inner layers, and these functions are respectively $[ReLU, ReLU, tanh]$. We follow the approach of \cite{svd_cl} for the experience selection (inclusion and removal of data pairs \cite{experience_selection}). In particular, we construct a matrix $TT\transp$, where $T$ consists of $p_{\max}$ labels, and compute its singular values. If the replacement of $i^{\text{th}}$ label by new label gives larger singular values than the old one, then the new data pair is added at the $i^{\text{th}}$ position of the replay buffer.
\par Our experimental results are depicted in Fig.\  \ref{fig:experiment2}. Due to the sudden change in $v_t$ at time instants shown by vertical grid lines, tube MPC has oscillations in roll angle. The performance of shallow MPC is affected at each instant of abrupt change, which depicts its incapability of generalization. However, deep MPC demonstrates a good generalization with only three hidden layers.   

\section{Conclusion}\label{s:epilogue}
A deep learning based algorithm is presented for safety critical systems by combining the approaches of adaptive control based label generation and tube MPC. 
%The proposed approach is advantageous over tube MPC when uncertainties are matched. 
A numerical experiment demonstrates that our approach with a single layer neural network (shallow MPC) outperforms tube MPC. The advantage of deep MPC is demonstrated in terms of further improvement in performance and convergence to a very close vicinity of origin. 
Future work may incorporate the results of \cite{deep_tube_learning, dynamic_tube_How, PDQ_intermittent, ref:PSC18}.

\clearpage

\acknowledgments{We gratefully acknowledge financial support from ONR MURI N00014-19-1-2373 and joint NSF CPS USDA grant 2018-67007-28379.}

\bibliography{MPC_Learning}  % .bib

\clearpage
\appendix

\section*{Appendix}

\section{Deep Neural Network}\label{s:adaptive}
Any continuous function $h$ on a compact set $\mathcal{X}$ can be approximated by a multi-layer network with number of layers $L \geq 2$ such that   
\begin{equation}
h(x) = W_L\transp \psi_L \left[ W_{L-1}\transp \psi_{L-1} \left[ \cdots \left [ \psi_1(x)\right] \right] \right] + \varepsilon^{\ast}(x),
\end{equation} 
where $x \in \mathcal{X}$, $\psi_i, W_i$ for $i=1, \ldots , L$, are activation functions and ideal weights, respectively, in the $i^{\text{th}}$ layer. The reconstruction error function $\varepsilon^\ast$ is bounded by a known constant $\bar{\varepsilon}^\ast > 0$ for each $x \in \mathcal{X}$, i.\ e.\ $\norm{\varepsilon^\ast(x)} \leq \bar{\varepsilon}^\ast$. Therefore, we can represent $h(\st_t)$ with the help of a neural network with a desired accuracy. If the neural network is not minimal then the ideal weights may not be unique. However, for the neural-adaptive controller design only the existence of ideal weights is assumed, which is always guaranteed when $h$ is a continuous function on a compact set \cite[\S 7.1]{Lewis_NN_99}. Let us define $\phi^{\ast}(x) \Let \psi_L \left[ W_{L-1}\transp \psi_{L-1} \left[ \cdots \left[ \psi_1(x)\right] \right] \right]$ as the output of the last activation layer under the ideal weights of hidden layers and $ W^{\ast} \Let W_{L} \in \R^{(n_L + 1) \times m} $ be the ideal weights of the output layer, then	 
\begin{equation}\label{e:NN_structure}
h(\st_t) = W^{\ast \top}\phi^{\ast}(\st_t) + \varepsilon^\ast(\st_t).
\end{equation} 
There are $n_L$ number of neurons in the output layer. The first row of $W^{\ast}$ represents the bias term and the first element of $\phi^\ast \in \R^{n_L+1}$ is $1$. The ideal hidden layer weights defining $\phi^{\ast}(\cdot)$ are neither known nor unique. 

\par
\begin{figure} 
	\centering
	\begin{adjustbox}{width = \columnwidth}
		\includegraphics{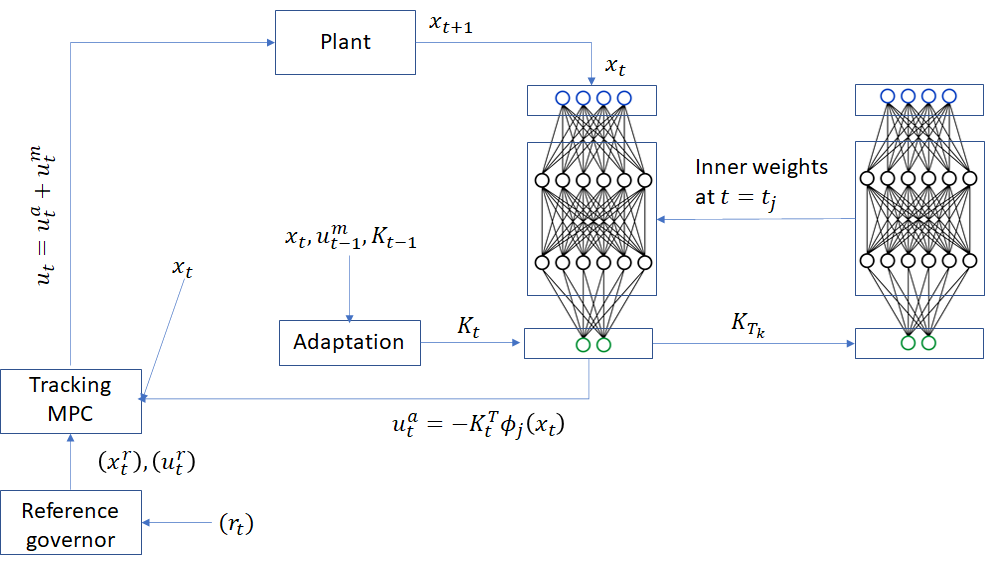}		
	\end{adjustbox}
	\caption{Schematic of DNN in the loop with MPC.}
	\label{fig:block_dia}
\end{figure}

 We update the weights of the output layer on the main machine in real time at each time instant with the help of a weight update law while keeping the weights of hidden layers fixed. The hidden layers are trained on a parallel secondary machine by using the approach \cite{DMRAC} in which the weights of the output layer are copied from the main machine at the start of the training and remain fixed during the training. Once the training of DNN on a secondary machine is complete, new weights of the hidden layers are updated on the main machine and remain fixed until new set of weights are again obtained from the secondary machine. The schematic of DNN in the loop with MPC is shown in Fig. \ref{fig:block_dia}. 
\begin{remark}
	\rm{
		For the implementation of our controller, we can access the output of the last activation layer of DNN without knowing the functions $\phi^\ast$ and $\varepsilon_{\phi_j}$.
	} 
\end{remark}
\begin{remark}[Necessity of second DNN]
	\rm{
		In many practical applications uncertainties appear in the dynamics through interaction with the environment and neural networks trained on one autonomous vehicle do not perform well on the other vehicle due to slight difference in hardware such as aperture of camera. In such situations deep learning based algorithms cannot be used for mass production without any provision of online training. The second DNN in Fig.\ \ref{fig:block_dia} allows to improve or re-adjust features with change in hardware or environment.   
	} 
\end{remark}
\par At time $t_0$, the neural network is initialized with random weights on both machines, and for a given $x$ as input, $\phi_0(x)$ denote the output of the last activation layer at $t_0$. Let $(t_j)_{j \in \N}$ denote the instants when the weights of hidden layers are updated on main machine after the completion of the $j^{\text{th}}$ training. Let $\phi_j(x)$ be the output of the last activation layer after the $j^{\text{th}}$ training for a given $x$ as input. We can use bounded neurons in the last activation layer, which results in bounded $\phi^\ast$ and $\phi_j$. Due to the universal approximation property of DNN, $\phi^\ast$ exists with bounded $\varepsilon^\ast$. We can assume that there exists $\varepsilon_{\phi_j}: \R^d \rightarrow \R^m$ for each $\phi_j$ such that $\phi^{\ast}(\st) = \phi_j(\st) + \varepsilon_{\phi_j}(\st)$ for each $\st \in \mathcal{X}$. The boundedness of both $\phi^\ast$ and $\phi_j$ ensures boundedness of $\varepsilon_{\phi_j}$. We need not to compute their bounds for the controller design. For $t \in \{t_j, t_j +1,  \ldots , t_{j+1}-1 \} $, \eqref{e:NN_structure} becomes  
\begin{equation}\label{e:NN_structure_general}
h(\st_t) = W^{\ast \top}\phi_j(\st_t) + \varepsilon_j(\st_t),
\end{equation} 
where $\varepsilon_j(\st_t) = \varepsilon^\ast(\st_t) + W^{\ast \top}\varepsilon_{\phi_j}(\st_t)$ is the overall reconstruction error. 

Notice that even when the weights of hidden layers are randomly assigned as in ELM \cite{ELM_06}, the universal approximation property of the neural network allows us to make the overall reconstruction error $\varepsilon_0(\cdot)$ as small as desired by increasing the width of the network. However, a network with trained hidden layers can capture several useful features, which in turn results in performance improvement \cite{DMRAC}. 

\par We employ 
\begin{equation}\label{e:adaptive_control}
\control_t^a = -K_t\transp \phi_j(\st_t)  
\end{equation}
as an adaptive (learning) control at time $t \in \{t_j, t_j +1,  \ldots , t_{j+1}-1 \} $, where $K_t$ is the weight of the output layer, which is trained according to the adaptive weight update law and $\phi_j$ is a feature basis function obtained from the last activation layer of DNN after $j^{\text{th}}$ training.
In the next subsections we provide the relevant details of the training of DNN. 

%	\subsection{Training of the outermost layer}\label{s:outer layer training}
\subsection{Adaptive learning of $W^\ast$ on the main machine}\label{s:outer layer training}
We make the following assumption:  
\begin{assumption}\label{as:bounds_uncertainty}
	\rm{
		There exist $\bar{W}_{i} > 0$ for $i = 1, \ldots , m$, and $\sigma, \bar{\varepsilon} > 0$ such that $\norm{W^{\ast (i)}} \leq \bar{W}_{i}$, for $i = 1, \ldots , m$, and $\norm{\phi_j(x)} \leq \sigma, \norm{\varepsilon_j(x)} \leq \bar{\varepsilon}$ for every $x \in \mathcal{X}$ and $j \in \Nz$.}   
\end{assumption}
The above assumption is standard in the literature \cite{HHN_08, NN_15, Joshi_DMRAC}. A priori knowledge about the bounds on the ideal weights $W^\ast$ of the output layer is useful to avoid parameter drift phenomenon. If the activation functions in the last hidden layer are bounded, i.\ e.\ sigmoidal, tanh, etc., then $\norm{\phi_j(x)}$ will also be bounded for each $j$ and for all $x \in \R^d$. 
\par We initialize $K_0$ such that $\norm{K_0^{(i)}} \leq \bar{W}_i$; $i = 1, \ldots , m$. For a given learning rate $0< \theta < 1$ and for $t \in \{t_j, t_j +1,  \ldots , t_{j+1}-1 \} $, we employ the following weight update law:
\begin{equation}\label{e:update_law}
\bar{K}_{t+1} = K_t + \frac{\theta}{\norm{\phi_j(\st_t)}^2} \phi_j(\st_t)\left( g(\st_t)^{\dagger}(\st_{t+1} - \bar{f}(\st_t, \control_t^m)) \right)\transp , 
\end{equation}
where $g(\st_t)^{\dagger} = \left( g(\st_t)\transp g(\st_t) \right)^{-1}g(\st_t)\transp $ represents the pseudo-inverse of the left invertible matrix $g(\st_t)$. Notice that first element in $\phi_j(\cdot)$ is one. Therefore, $\norm{\phi_j(x)}^2 \geq 1$ for all $x \in \R^d$ and $j \in \Nz$, which avoids any possibility of division by zero. 
\par We employ the discrete projection method to ensure boundedness of $K_t^{(i)}$ for $i= 1, \ldots , m$, as follows:
\begin{equation}\label{e:projection}
K_t^{(i)} = \proj \bar{K}_t^{(i)} = \begin{cases} \bar{K}_t^{(i)} & \text{ if } \norm{\bar{K}_t^{(i)}} \leq \bar{W}_{i} \\
\frac{\bar{W}_{i}}{\norm{\bar{K}_t^{(i)}}} \bar{K}_t^{(i)} & \text{ otherwise. } \end{cases}
\end{equation}
Let $\tilde{K}_t \Let K_t - W^{\ast}$ and $\tilde{\control}_t \Let \control_t^a + h(\st_t) = - \tilde{K}_t \transp \phi_j(\st_t) + \varepsilon_j(\st_t) $. 
It is evident that $\norm{K_t}_F^2 = \trace(K_t\transp K_t) = \sum_{i=1}^m \norm{K_t^{(i)}}^2 \leq \sum_{i=1}^m \bar{W}_i^2 \teL \bar{W}$ for all $t$ due to the projection. 
Therefore, the neuro-adaptive control component $\control_t^a $ is bounded, i.\ e.\
\begin{equation*}
%\begin{aligned}
\norm{\control_t^a} = \norm{K_t\transp \phi_j(\st_t)} \leq \norm{K_t}\norm{\phi_j(\st_t)} \leq \norm{K_t}_F \sigma \\
\leq \sqrt{\bar{W}}\sigma \teL \authority^a,
%\end{aligned}
\end{equation*}
for $t \in \{t_j, t_j +1,  \ldots , t_{j+1}-1 \} $ and for all $t_j$.
The apparent disturbance term $g(\st_t) \left( \control_t^a + h(\st_t)\right)$ in \eqref{e:actual_agent} is also bounded, i.\ e.\ 
\begin{equation}\label{e:overall_disturbance_bound}
\begin{aligned}
& \norm{g(\st_t) \left( \control_t^a + h(\st_t)\right)} \leq \norm{g(\st_t){\control}_t^a} + \norm{g(\st_t)h(\st_t)} \\
& \quad \leq \delta_g\norm{\control_t^a} + w_{\max} \leq \delta_g \authority^a + w_{\max} \teL w_{\max}^{\prime}.
\end{aligned}
\end{equation}    	

\subsection{Self-supervised learning of $\phi^\ast$ on a secondary machine}\label{s:inner layer training}
Let $(T_k)_{k \in \N}$ represent time instants when we begin the $k^{\text{th}}$ training of DNN. Let $p_0$ data samples are required for the training, which are stored in a buffer of size $p_{\max} > p_0$. We do not have access of the labeled data pairs $(x, \phi^\ast(x))$. Therefore, we follow an approach similar to that of \cite{DMRAC} for the data collection and training.
\par We fix $T_1 \geq p_0$ and for each $t \leq T_1$, the labeled pairs $(\st_t,\control_t^a)$ are stored in the buffer. Recall that $\control_t^a = -K_t\transp \phi_0(\st_t)$ for $t \leq T_1 < t_1$, where $\phi_0(\cdot)$ is obtained by the random initialization of the weights of hidden layers. At $t=T_1$, we randomly sample $p_0$ data pairs for the training of DNN. We fix the weights of the output layer to be $-K_{T_1}$ and train the network. Notice that the training of DNN does not affect the operation of system because the controlled system still employs $\control_t^a = -K_t\transp \phi_0(\st_t)$ as the adaptive control in which only $K_t$ is updated at each time instant by using the weight update law discussed in \secref{s:adaptive}. At $t = t_1$, we get our first trained network. For $t \in \{t_1, \ldots , t_2-1 \}$, we employ $\control_t^a = -K_t\transp \phi_1(\st_t)$ as an adaptive control. This process of training, exploiting and storing is repeated at each time $t$. For each $k \in \N$, $W_L$ is set to be $-K_{T_k}$ in the secondary DNN and remain fixed during the training. Therefore, we are interested in finding the weights $W_{1:L-1} \Let W_1, \ldots, W_{L-1}$ which minimize the following cost for a given input $x_t$ and corresponding label $\control_t^a$:
\begin{equation*}
\ell\left( (x_t, \control_t^a), W_{1:L-1} \right) \Let \norm{\control_t^a + K_{T_k}\transp \psi_L \left[ W_{L-1}\transp \psi_{L-1} \left[ \cdots \left [ \psi_1 (x_t)\right] \right] \right]}^2. 
\end{equation*}
Let $\mathcal{D}_k \Let \left( x^i,u^i \right)_{i=1}^{p_0}$ be training data consisting of $p_0$ data points randomly sampled from  the buffer for the $k^{\text{th}}$ training. The following loss function is considered for the training of DNN:
\begin{equation*}
\mathcal{L}(\mathcal{D}_k, W_{1:L-1}) = \frac{1}{p_0} \sum_{i=1}^{p_0} \ell\left( x^i, W_{1:L-1} \right).
\end{equation*}  
\par At $t = p_{\max}-1$, the buffer becomes full. So new data can be added after the removal of some old data by using some suitable experience selection method \cite{experience_selection}. The available approaches are based on retaining the most informative data based on some criterion \cite{sparsification_data} and ensuring sufficient diversity \cite{diverse_data}. Our present approach is compatible with any existing method of experience selection. However, different methods may result in different performance for different problems and their choice may also depend on the availability of resources. We keep the method of experience selection open for the choice of users.

\section{Stability}

\par We recall the following definition:
\begin{definition}[\cite{Ioannou_Fidan_tutorial}, page 117]\label{def:small_disturbance}
	\rm{
		The vector sequence $(s_t)_{t \in \Nz}$ is called $\mu$ small in mean square sense if it satisfies $\sum_{t=k}^{k+N-1}\norm{s_t}^2 \leq Nc_0 \mu + c_0^{\prime}$ for all $k \in \N$, a given constant $\mu \geq 0$ and some $N \in \N$, where $c_0, c_0^{\prime} \geq 0$. } 
\end{definition}
Some straightforward arguments as in \cite[\S 4.11.3]{Ioannou_Fidan_tutorial} give us the following result:

\begin{lemma}\label{lem:adaptation}
	Consider the dynamical system \eqref{e:system}, weight update law \eqref{e:update_law} and the projection method \eqref{e:projection}. Let the Assumption \ref{as:bounds_uncertainty} hold and define $V_a(K_t) \Let \frac{1}{\theta}\trace(\tilde{K}_t\transp\tilde{K}_t)$. Then for all $t$,
	\begin{enumerate}[label={(\rm \roman*)}, leftmargin=*, widest=3, align=left, start=1]
		\item \label{lem:bound_Va} $V_a(K_t) \leq \frac{4}{\theta} \bar{W}$, 
		\item \label{lem:drift_Va} $V_a(K_{t+1}) - V_a(K_t)  \leq - \frac{1- \theta}{\sigma^2} \norm{\tilde{\control}_t}^2 +  \norm{\varepsilon(\st_t)}^2$,
		\item \label{lem:small_tildeu} $\tilde{\control}_t$ is $\bar{\varepsilon}^2$ small in mean square sense with $c_0 = \frac{\sigma^2}{1-\theta }$ and $c_0^{\prime} = \frac{4c_0}{\theta}\bar{W}$ as per the Definition \ref{def:small_disturbance}.	
	\end{enumerate}			
\end{lemma}
We provide a proof of Lemma \ref{lem:adaptation} in the appendix. Let $X_c(\st_t^r)$ be the level set around $\st_t^r$ of radius $c$ generated by $V_m(\st_t)$ and $X_c$ be their union. In particular,	
\begin{equation}\label{e:tube}
\begin{aligned}
X_c(\st_t^r) & \Let \{x_t \in \R^d \mid V_m(x_t) \leq c; \; c > 0  \}, \\
X_c & \Let \cup_{t=0}^{N} X_c(\st_t^r) .
\end{aligned}
\end{equation}
Properties of the value function are summarized in the following Lemma. These results are standard in the literature \cite{ref:rawlings-09}. We provide their proofs in the appendix for completeness. 
\begin{lemma}\label{lem:value_mpc}
	\begin{enumerate}[label={(\rm \roman*)}, leftmargin=*, widest=3, align=left, start=1]
		\item \label{e:satisfaction_terminal_set} If $\alpha \geq c$ then $\st_{t+N\mid t} \in \mathcal{X}_f$ for every $\st_t \in X_c(\st_t^r) $. 
		\item \label{e:iss_Lyapunov1} \cite[Lemma 3]{MWGC} There exist $c_2 > c_1 > 0$ such that 
		\begin{align*} 
		& c_1 \norm{\st_t -\st_t^r}^2 \leq \costps(\st_t, \control_{t\mid t}^\ast) \leq V_m(\st_t) \leq c_2 \norm{\st_t -\st_t^r}^2 .
		\end{align*}
		%			\item \label{e:Lipscitz_continuity_value_function} \cite[Proposition 3]{Mayne_tube_NLMPC} There exists $c_3> 0$ such that 
		%			\begin{equation*}
		%			\abs{V_m(x) - V_m(y)} \leq c_3 \norm{x-y} \text{ for every } x,y \in X_c + \mathds{W}^\prime 
		%			\end{equation*}
	\end{enumerate}
\end{lemma} 
Lemma \ref{lem:value_mpc}-\ref{e:satisfaction_terminal_set} ensures the satisfaction of terminal constraint on states just by construction. Refer to \cite[Proposition 1]{Stewart_Rawling_11} and \cite[Proposition 1]{Mayne_tube_NLMPC} for minor differences due to \eqref{e:system}, \eqref{e:MPC} and Assumption \ref{as:stability}. For the purpose of analysis, we define an intermediate optimization problem by replacing $\st_t$ in \eqref{e:constraint_initial} by $\st_{t\mid t-1}$. In particular,
\begin{equation}\label{e:intermediate_problem}
\hat{V}_m(\st_{t\mid t-1}) \Let \min_{(\control_{t+i \mid t})_{i=0}^{N-1}} \left\{ \eqref{e:cost_function} \mid \st_{t \mid t} = \st_{t \mid t-1}, \eqref{e:constraint_first_control}, \eqref{e:constraint_dynamics}, \eqref{e:constraint_remaining_control} \right\}.
\end{equation} 
Notice that we keep $\control_t^a = -K_t\transp\phi_j(\st_t)$ fixed in both problems \eqref{e:MPC} and \eqref{e:intermediate_problem}, respectively, and therefore, we can follow the following convention:
\begin{equation}\label{e:convention}
\hat{V}_m(\st_{t\mid t-1}) \leq c \implies \st_{t\mid t-1} \in \mathcal{X}_c(\st_t^r). 
\end{equation}
\begin{remark}
	\rm{
		Notice that the constraint on the first control \eqref{e:constraint_first_control} includes $\control_t^a$ to make MPC aware of the adaptive action. Since $\control_t^a = -K_t\transp \phi_j(\st_t)$ nonlinealry depends on $\st_t$ due to the nonlinear function $\phi_j$, the set-valued control move map becomes state-dependent. Our analysis is based on using the value function of MPC \eqref{e:MPC} as a candidate Lyapunov function. The presence of state-dependent constraint \eqref{e:constraint_first_control} prohibits us to prove robustness of MPC by invoking \cite[propositions 7,8 or 11]{examples_nonrobust}. We defined an intermediate optimization problem \eqref{e:intermediate_problem} to get rid of the above difficulty. Due to the above-mentioned technical difficulty the results of \cite[propositions 2 and 4]{Mayne_tube_NLMPC} are not directly applicable here. 
	}
\end{remark}
Important results related to tube MPC are summarized in the following Lemma. Refer to \cite[Proposition 2, Proposition 4]{Mayne_tube_NLMPC} for a detailed discussion. We provide their proofs in the appendix to highlight the adjustments and for completeness.
\begin{lemma}\label{lem:stability}
	If Assumption \ref{as:stability} is satisfied, then for all $t$ for every $\st_t \in X_c(\st_t^r)$ the following hold:
	\begin{enumerate}[label={(\rm \roman*)}, leftmargin=*, widest=3, align=left, start=1]
		\item \label{e:bound_on_predicted_next} $\hat{V}_m(\st_{t+1 \mid t}) - V_m(\st_t) \leq -\costps(\st_t, \control_{t \mid t}^\ast)$, and $\st_{t+1\mid t} \in \mathcal{X}_c$.
		\item \label{e:inclusion_next_state} $\st_{t+1} \in \mathcal{X}_c(\st_{t+1}^r) + \mathds{W}^\prime$.
		\item \label{e:actual_predicted} $V_m(\st_{t+1}) - \hat{V}_m(\st_{t+1 \mid t}) \leq c_3 \norm{g(\st_t)\tilde{\control}_t}$.
		\item \label{e:ISS}There exists $\gamma < 1$ such that
		\begin{equation*}
		V_m(\st_{t+1}) \leq \gamma V_m(\st_t) +  c_3 \norm{g(\st_t)\tilde{\control}_t}.
		\end{equation*}
	\end{enumerate}
\end{lemma}
The Lemma \ref{lem:stability}-\ref{e:ISS} along with Lemma \ref{lem:value_mpc}-\ref{e:iss_Lyapunov1} ensures that the controlled system is input-to-state stable (ISS) because it admits $V_m$ as an ISS Lyapunov function \cite[Lemma 3.5]{ISS_discrete}. In the case of structured uncertainty $\norm{\tilde{\control}_t} \rightarrow 0$ as $t \rightarrow  \infty$, which implies $\norm{\st_t} \rightarrow 0$ \cite[Theorem 1]{MWGC}. Such results are not available in the presence of unstructured uncertainty. However, the existence of invariant and attractive tubes is possible when $\wnoise_{\max}$ and $\authority^a$ are small. We have the following result:  
\begin{proposition}\label{prop:main_tube}
	Let us define $\bar{c} \Let \frac{c_2 c_3 }{c_1}(\delta_g\authority^a + \wnoise_{\max})$.
	If $\delta_g\authority^a + \wnoise_{\max}  < \frac{c_1}{c_2 c_3}c$, then for all $t\geq N$, the following hold:
	\begin{enumerate}[label={(\rm \roman*)}, leftmargin=*, widest=3, align=left, start=1]
		\item \label{prop:invariant_tube} for every $\st_t \in \mathcal{X}_c(0)\setminus \mathcal{X}_{\bar{c}}(0)$, $V_m(\st_{t+1}) < V_m(\st_t)$,
		\item for every $\st_t \in \mathcal{X}_{\bar{c}}(0)$, $\st_{t+1} \in \mathcal{X}_{\bar{c}}(0)$.
		\item In addition, if $\mathcal{X}_c \subset \mathcal{X}$, then $\st_t \in \mathcal{X}$ for all $t$.
	\end{enumerate}
\end{proposition}
The Proposition \ref{prop:main_tube} has similar arguments as in \cite[Proposition 4]{Mayne_tube_NLMPC} and confirms the existence of an invariant tube $\mathcal{X}_c(0)$ and an attractive tube $\mathcal{X}_{\bar{c}}(0) \subset \mathcal{X}_c(0)$.  
%	   \todo[inline]{invariant and attractive}
\par Suppose there exists some $\hat{N} \geq N$ and $\hat{c}$ such that $\st_{\hat{N}} \in \mathcal{X}_{\hat{c} }(0) \Let \{ x \mid V_m(x) \leq \hat{c} \leq c \}$. Since $c_3$ is a Lipschitz constant of $V_m$ on a compact set $\mathcal{X}_c + \mathds{W}^\prime \supset \mathcal{X}_c(0) \supset \mathcal{X}_{\hat{c}}(0)$, there exists $\hat{c}_3$, which satisfies Lemma \ref{lem:stability}-\ref{e:ISS}. Similarly, let there exist $\hat{\delta}_g \leq \delta_g$ such that $\norm{g(x)} \leq \hat{\delta}_g$ for every $x \in \mathcal{X}_{\hat{c}}(0)$. Since $\bar{c}$ depends on $c_3$ and $\delta_g$, their reduction will result in shrinkage of the attractive tube $\mathcal{X}_{\bar{c}}(0)$. Moreover, since any level set within $\mathcal{X}_{\bar{c}}(0)$ is invariant due to Proposition \ref{prop:main_tube}-\ref{prop:invariant_tube}, a further shrinkage is possible. However, asymptotic convergence is still not guaranteed. If $\gamma^2 < \frac{1}{2}$, then we can get a stronger result provided a certain condition in terms of $c_3$ and $\delta_g$ is satisfied, and the reconstruction error $\varepsilon$ has small gain type property within the invariant tube. We make the following assumption:
\begin{assumption}\label{as:notation_small_gain}
	There exists $\beta > 0$ such that $\norm{\varepsilon_j(\st)} \leq \beta \norm{\st}^2$ for all $\st \in \mathcal{X}_{\hat{c}}(0)$ and $j \in \Nz$. 	 
\end{assumption}
Generally, the norm bound on the reconstruction error is assumed to be linear in $\norm{x}$\cite{HHN_08}. We assumed it to be quadtratic, otherwise the above assumption is standard in literature. 
We have the following result:
\begin{theorem}\label{th:asymptotic}
	Consider the dynamical system \eqref{e:system} controlled by the Deep MPC, and let assumptions \ref{as:bounds_uncertainty}, \ref{as:stability} and \ref{as:notation_small_gain} hold. If $\wnoise_{\max}^\prime < \frac{c_1}{c_2 c_3}c$, $\gamma^2 < \frac{1}{2}$ and $\beta < \frac{c_1 m}{\sqrt{2}\sigma \hat{c}_3 \hat{\delta}_g}\sqrt{(1-2\gamma^2)(1-\theta)}$, 
	then $\norm{\st_t} \rightarrow 0$ as $t \rightarrow \infty$.
\end{theorem}
Notice that the main results of tube-based MPC (Proposition \ref{prop:main_tube}) are valid for small disturbances. The Theorem \ref{th:asymptotic} extends Proposition \ref{prop:main_tube} by guaranteeing convergence of states to origin under the conditions on $\gamma$ and $\beta$. Smaller value of $\gamma$ refers to the faster convergence of the value function of nominal MPC. Generally, reconstruction error is comparatively very small with respect to the disturbance. Therefore, the condition on $\gamma$ and $\beta$ are reasonable, and they can be verified in both theoretical and empirical manner.    	    

\section{Proofs}
\begin{proof}[Proof of Lemma \ref{lem:adaptation}]
	\begin{enumerate}[label={(\rm \roman*)}, leftmargin=*, widest=3, align=left, start=1]
		\item Since $V_a(K_t) = \frac{1}{\theta} \trace(\tilde{K}_t\transp \tilde{K}_t) = \frac{1}{\theta} \sum_{i=1}^m \norm{K_t^{(i)} - W^{\ast (i)}}^2 \leq \frac{4}{\theta} \sum_{i=1}^m \bar{W}_i^2 = \frac{4}{\theta} \bar{W}$.
		\item We first compute
		\begin{align*}
		V_a(K_{t+1}) &= \frac{1}{\theta} \trace(\tilde{K}_{t+1}\transp \tilde{K}_{t+1}) \\
		&= \frac{1}{\theta}\trace \left( (\bar{K}_{t+1} - W^\ast)\transp (\bar{K}_{t+1} - W^\ast)\right) + \frac{1}{\theta}\trace(\alpha_t),
		\end{align*}
		where 
		\begin{equation*}
		\begin{aligned}
		\alpha_t &= (K_{t+1} - \bar{K}_{t+1})\transp (K_{t+1} - \bar{K}_{t+1}) 
		 + 2 (K_{t+1} - \bar{K}_{t+1})\transp (\bar{K}_{t+1} - W^{\ast})\\
		&= -(K_{t+1} - \bar{K}_{t+1})\transp (K_{t+1} - \bar{K}_{t+1}) + 2(K_{t+1} - \bar{K}_{t+1})\transp (K_{t+1}-W^\ast). 
		\end{aligned}
		\end{equation*}
		One important property of the projection \eqref{e:projection} is the following \cite[(4.61)]{Ioannou_Fidan_tutorial}:%, \cite[page 341]{Landau_Karimi_11}:
		\begin{equation}\label{e:effect_projection}
		(W^{\ast (i)} - K_t^{(i)})\transp (\bar{K}_{t}^{(i)} - K_{t}^{(i)})  \leq 0 \text{ for each } i = 1, \ldots , m.
		\end{equation}
		Since $(K_{t+1}^{(i)} - \bar{K}_{t+1}^{(i)})\transp (K_{t+1}^{(i)}-W^\ast) \leq 0$ due to \eqref{e:effect_projection}, we can ensure $\trace(\alpha_t) \leq 0$. Therefore,
		\begin{align*}
		V_a(K_{t+1}) & \leq \frac{1}{\theta}\trace \left( (\bar{K}_{t+1} - W^\ast)\transp (\bar{K}_{t+1} - W^\ast)\right) \\
		&= V_a(K_t) + \frac{1}{\norm{\phi_j(\st_t)}^2} \trace \left( \tilde{\control}_t \tilde{\control}_t\transp + 2 \tilde{K}_t \transp \phi_j(\st_t) \tilde{\control}_t\transp \right).
		\end{align*}
		By substituting $\tilde{K}_t \transp \phi_j(\st_t) = -\tilde{\control}_t + \varepsilon_j(\st_t) $ in the above inequality, we get
		\begin{align*}
		V_a(K_{t+1}) & \leq V_a(K_t) + \frac{1}{\norm{\phi_j(\st_t)}^2} \trace \left( \theta \tilde{\control}_t \tilde{\control}_t\transp + 2 (-\tilde{\control}_t + \varepsilon_j(\st_t)) \tilde{\control}_t\transp \right) \\
		&= V_a(K_t) + \frac{1}{\norm{\phi_j(\st_t)}^2} \left( \theta \norm{\tilde{\control}_t}^2  - 2 \tilde{\control}_t\transp(\tilde{\control}_t - \varepsilon_j(\st_t))  \right) \\
		& = V_a(K_t) + \frac{1}{\norm{\phi_j(\st_t)}^2} \left( (\theta -2) \norm{\tilde{\control}_t}^2  + 2 \tilde{\control}_t\transp \varepsilon_j(\st_t)  \right) \\
		& \leq V_a(K_t) + \frac{1}{\norm{\phi_j(\st_t)}^2} \left( (\theta -1) \norm{\tilde{\control}_t}^2  + \norm{\varepsilon_j(\st_t)}^2  \right) \\
		& = V_a(K_t) - \frac{1- \theta}{\norm{\phi_j(\st_t)}^2} \norm{\tilde{\control}_t}^2 + \frac{1}{\norm{\phi_j(\st_t)}^2} \norm{\varepsilon_j(\st_t)}^2 \\
		& \leq V_a(K_t) - \frac{1- \theta}{\sigma^2} \norm{\tilde{\control}_t}^2 +  \frac{1}{m^2}\norm{\varepsilon_j(\st_t)}^2, \\
		\end{align*}	
		where the last inequality is due to $m^2 \leq \norm{\phi_j(\st_t)}^2 \leq \sigma^2$. Therefore,
		\begin{equation*}
		V_a(K_{t+1}) - V_a(K_t)  \leq - \frac{1- \theta}{\sigma^2} \norm{\tilde{\control}_t}^2 +  \frac{1}{m^2}\norm{\varepsilon_j(\st_t)}^2 .
		\end{equation*}
		\item Consider Lemma \ref{lem:adaptation}-\ref{lem:drift_Va} to get
		\begin{align*}
		\frac{1 - \theta }{\sigma^2 }\norm{\tilde{\control}_t}^2 & \leq - V_a(K_{t+1}) + V_a(K_t) + \frac{1}{m^2} \norm{\varepsilon_j(\st_t)}^2 \\
		& \leq - V_a(K_{t+1}) + V_a(K_t) + \frac{\bar{\varepsilon}^2}{m^2}.
		\end{align*}
		By summing from $t=k$ to $k+N-1$ in both sides, we get
		\begin{equation*}
		\begin{aligned}
		&\frac{1 - \theta }{\sigma^2 } \sum_{t=k}^{k+N-1} \norm{\tilde{\control}_t}^2 \leq V_a(K_k) + \frac{N}{m^2} \bar{\varepsilon}^2 \leq \frac{4}{\theta} \bar{W} + \frac{N}{m^2} \bar{\varepsilon}^2.  
		\end{aligned}
		\end{equation*}
		Therefore, $\tilde{\control}_t$ is $\bar{\varepsilon}^2$ small in mean square sense with $c_0 = \frac{\sigma^2}{(1-\theta)m^2 }$ and $c_0^{\prime} = \frac{4c_0}{\theta}m^2\bar{W}$ as per the Definition \ref{def:small_disturbance}.	
	\end{enumerate}			 	
\end{proof}	
%%%%%%%%%%%%%%%%%%%%%%%%%	
\begin{proof}[Proof of Lemma \ref{lem:value_mpc}]
	\begin{enumerate}[label={(\rm \roman*)}, leftmargin=*, widest=3, align=left, start=1]
		\item We recall the definitions of $\mathcal{X}_c(\st_t^r)$ and $\mathcal{X}_f$ from \eqref{e:terminal_set} and \eqref{e:tube}, respectively. Now, it is immediate to notice that $\st_t \in X_c(\st_t^r) \implies V_m(\st_t) \leq c \implies \costfinal(\st_{t+N \mid t}) \leq V_m(\st_t) \leq c \leq \alpha  \implies \st_{t+N \mid t} \in \mathcal{X}_f$.
		\item Since $Q \succ 0$ and $f,g$ are Lipschitz continuous, by \cite[Lemma 3]{MWGC} there exist $c_1, c_2 > 0$ such that
		Lemma \ref{lem:value_mpc}-\ref{e:iss_Lyapunov1} hold. We mention key steps here for completeness. Since $ V_m(\st_t) \geq \costps(\st_t, \control_t^m) \geq \norm{\st_t -\st_t^r}^2_Q$, we can choose $c_1 = \lambda_{\min}(Q)$. 
		\par Let $f, g$ be Lipschitz continuous with Lipschitz constants $L_f$ and $L_g$, respectively. We can notice that \eqref{e:MPC} has no constraints on states and the constraints on control can be satisfied by $(\control_i^r)_{i=t}^{t+N-1}$ at time $t$. 
		\par Let us recall the definition of the cost function \eqref{e:cost_function}, then due to the optimality of $V_m(\st_t)$, we get
		\begin{align*}
		V_m(\st_t) &\leq  V(\st_t, (\control_{t+i}^r)_{i=0}^{N-1}) \\
		& = \st_{t+N\mid t}\transp Q_f \st_{t+N\mid t} + \sum_{i=0}^{N-1} (\st_{t+i \mid t}-\st_{t+i}^r)\transp Q (\st_{t+i \mid t}-\st_{t+i}^r) \\
		& \leq \lambda_{\max}(Q_f)\norm{\st_{t+N\mid t}}^2 + \sum_{i=0}^{N-1} \lambda_{\max}(Q) \norm{\st_{t+i\mid t} - \st_{t+i}^r}^2 .
		\end{align*}
		The above inequality is due to the substitution $ (\control_i)_{i=t}^{t+N-1}= (\control_i^r)_{i=t}^{t+N-1}$. Further,
		\begin{align*}
		\st_{t+i \mid t} - \st_{t+i}^r &= f(\st_{t+i -1 \mid t}) -f(\st_{t+i -1}^r) \\
		& \quad + \left( g(\st_{t+i -1 \mid t})-g(\st_{t+i -1}^r)\right) \control_{t+i-1}^r \\
		\norm{\st_{t+i \mid t} - \st_{t+i}^r} & \leq \left( L_f+L_g\norm{\control_{t+i-1}^r} \right) \norm{\st_{t+i-1 \mid t} - \st_{t+i-1}^r}  \\
		&\leq \bar{L}\norm{\st_{t+i-1 \mid t} - \st_{t+i-1}^r}  \leq \bar{L}^i\norm{\st_{t} - \st_{t}^r},
		\end{align*}
		where $\bar{L} = L_f+L_g\authority^r$. Since $\st_{t+N}^r = \zeros$ for all $t$, there exists $c_2 = \bar{L}^N\lambda_{\max}(Q_f) + \sum_{i=0}^{N-1}\bar{L}^i\lambda_{\max}(Q) > \lambda_{\min}(Q) = c_1$.
	\end{enumerate}
\end{proof}
%%%%%%%%%%%%%%%%%%%

%%%%%%%%%%%%%%%
\begin{proof}[Proof of Lemma \ref{lem:stability}]
	\begin{enumerate}[label={(\rm \roman*)}, leftmargin=*, widest=3, align=left, start=1]
		\item Since $\control_{t+1 \mid t}^\ast \in \controlset^\prime$, we get $\control_{t+1 \mid t}^\ast + \control_{t+1}^a \in \controlset$. Therefore, $\control_{t+1\mid t+1} = \control_{t+1 \mid t}^\ast $ is feasible for \eqref{e:intermediate_problem} at time $t+1$. Since $\control_{t+i+1 \mid t}^\ast \in \controlset^\prime$, for $i=1, \ldots, N-2$ the control sequence $\control_{t+i+1 \mid t+1} = \control_{t+i+1 \mid t}^\ast$ is also feasible at time $t+1$ for \eqref{e:intermediate_problem}. Under the above control sequence $\st_{t+N\mid t+1} = \st_{t+N \mid t} \in \mathcal{X}_f$ because $\st_{t+1\mid t+1} = \st_{t+1\mid t}$ in \eqref{e:intermediate_problem}. Therefore, $\control_{t+N \mid t+1} = \control^\prime$ is feasible for some $\control^\prime \in \controlset^\prime$ satisfying the Assumption \ref{as:stability}. In this way, we have constructed a feasible control sequence $(\control_{t+i+1\mid t+1})_{i=0}^{N-1}$ for \eqref{e:intermediate_problem} and due to the optimality of $\hat{V}_m(\st_{t+1\mid t})$, by substituting the feasible control sequence $(\control_{t+i+1\mid t+1})_{i=0}^{N-1} = \{u^\prime , (\control_{t+i+1 \mid t}^\ast)_{i=0}^{N-2} \}$ in \eqref{e:intermediate_problem}, we get 
		\begin{align*}
		& \hat{V}_m(\st_{t+1 \mid t}) \leq \costfinal(\st_{t+N+1 \mid t+1}) + \sum_{i=0}^{N-1}\costps(\st_{t+1+i \mid t+1}, \control_{t+1+i \mid t+1}) \\
		& = \costfinal(\st_{t+N +1 \mid t+1}) + \costps(\st_{t+N \mid t+1}, \control_{t+N \mid t+1})  + \sum_{i=0}^{N-2}\costps(\st_{t+1+i \mid t+1}, \control_{t+1+i \mid t+1}) \\
		& = \costfinal(\st_{t+N + 1 \mid t+1}) + \costps(\st_{t+N \mid t}, \control_{t+N \mid t+1}) + \sum_{i=0}^{N-2}\costps(\st_{t+1+i \mid t}, \control_{t+1+i \mid t}^\ast) \\
		& = \costfinal(\st_{t+N +1 \mid t+1}) + \costps(\st_{t+N \mid t}, \control^\prime) + \sum_{i=1}^{N-1}\costps(\st_{t+i \mid t}, \control_{t+i \mid t}^\ast) \\
		& = \costfinal(\st_{t+N+1 \mid t+1}) + \costps(\st_{t+N \mid t}, \control^\prime) - \costfinal(\st_{t+N \mid t}) - \costps(\st_{t}, \control_{t\mid t}^\ast) + V_m(\st_t) \\
		& = \costfinal(\bar{f}(\st_{t+N \mid t}, \control^\prime)) + \costps(\st_{t+N \mid t}, \control^\prime) - \costfinal(\st_{t+N \mid t}) - \costps(\st_{t}, \control_{t\mid t}^\ast) + V_m(\st_t) \\
		& \leq  - \costps(\st_{t}, \control_{t\mid t}^\ast) + V_m(\st_t) 
		\end{align*}  
		due to the Assumption \ref{as:stability}. Therefore, $\hat{V}_m(\st_{t+1 \mid t}) \leq V_m(\st_t) \leq c$, which implies $\st_{t+1 \mid t} \in \mathcal{X}_c(\st_{t+1}^r) \subset \mathcal{X}_c$ due to our convention \eqref{e:convention}.
		\item Since $\st_{t+1\mid t} = \bar{f}(\st_t, \control_t^m) \in \mathcal{X}_c(\st_{t+1}^r)$, we get $\st_{t+1} = \st_{t+1\mid t} + g(\st_t)\tilde{\control}_t \in \mathcal{X}_c(\st_{t+1}^r) + \mathds{W}^\prime \subset \mathcal{X}_c + \mathds{W}^\prime$ due to the Lemma \ref{lem:stability}-\ref{e:bound_on_predicted_next}.
		\item  Notice that the optimization problems \eqref{e:MPC} and \eqref{e:intermediate_problem} do not have constraints on state. Let $(v_{t+i+1})_{i=0}^{N-1}$ be the minimizer of \eqref{e:intermediate_problem} at $t+1$, which means $\hat{V}_m(\st_{t+1\mid t}) = V_m(\st_{t+1\mid t}, (v_{t+i+1})_{i=0}^{N-1})$. Since $(v_{t+i+1})_{i=0}^{N-1}$ satisfies constraints on control \eqref{e:constraint_first_control} and \eqref{e:constraint_remaining_control}, it is feasible for \eqref{e:MPC} at $t+1$. Therefore, due to the optimality of $V_m(\st_{t+1})$, we get
		\begin{align*}
		&V_m(\st_{t+1}) \leq V_m(\st_{t+1}, (v_{t+i+1})_{i=0}^{N-1}), 
		\end{align*}
		which in turn implies 
		\begin{align*}
		V_m(\st_{t+1}) - \hat{V}_m(\st_{t+1\mid t}) & \leq V_m(\st_{t+1}, (v_{t+i+1})_{i=0}^{N-1}) - V_m(\st_{t+1\mid t}, (v_{t+i+1})_{i=0}^{N-1}) \\
		& \leq \abs{V_m(\st_{t+1}, (v_{t+i+1})_{i=0}^{N-1}) - V_m(\st_{t+1\mid t}, (v_{t+i+1})_{i=0}^{N-1}) }. 
		\end{align*}
		Now we notice that the cost function \eqref{e:cost_function} is Lipschitz continuous in its first argument on the set $\mathcal{X}_c + \mathds{W}^\prime$ while keeping the second argument fixed and $\st_{t+1}, \st_{t+1\mid t} \in \mathcal{X}_c + \mathds{W}^\prime$. Since $v_{t+i+1} \in \controlset$ for $i = 0, \ldots, N-1$, there exists some $c_3 > 0$ such that 
		\begin{align*}
		& \abs{V_m(\st_{t+1}, (v_{t+i + 1})_{i=0}^{N-1}) - V_m(\st_{t+1\mid t}, (v_{t+i+1})_{i=0}^{N-1}) }	\\
		& \quad \leq c_3 \norm{\st_{t+1}- \st_{t+1\mid t}} = c_3 \norm{g(\st_t)\tilde{\control}_t}.
		\end{align*}
		Since $t$ was arbitrary, the above result holds for all $t$.
		\item We compute a bound on $V_m(\st_{t+1})- V_m(\st_t) = V_m(\st_{t+1}) - \hat{V}_m(\st_{t+1 \mid t}) + \hat{V}_m(\st_{t+1 \mid t})- V_m(\st_{t}) $. Then by combining the results of Lemma \ref{lem:stability}-\ref{e:bound_on_predicted_next} and Lemma \ref{lem:stability}-\ref{e:actual_predicted}, we get $V_m(\st_{t+1})- V_m(\st_t) \leq -\costps(\st_t, \control_{t \mid t}^\ast) + c_3 \norm{g(\st_t)\tilde{\control}_t}$. Then due to Lemma \ref{lem:value_mpc}-\ref{e:iss_Lyapunov1}, we have  
		\begin{align*}
		V_m(\st_{t+1}) \leq \gamma V_m(\st_t) +  c_3 \norm{g(\st_t)\tilde{\control}_t},
		\end{align*} 
		where $\gamma = 1-\frac{c_1}{c_2}< 1$.		
	\end{enumerate}
\end{proof}
%%%%%%%%%%%%%%%%

%%%%%%%%%%%%%%%%
\begin{proof}[Proof of Proposition \ref{prop:main_tube}]
	\begin{enumerate}[label={(\rm \roman*)}, leftmargin=*, widest=3, align=left, start=1]
		\item We can observe that $c_3\norm{g(\st_t) \tilde{\control}_t} \leq c_3\wnoise_{\max}^{\prime} = \frac{c_1}{c_2}\bar{c}$. Therefore, due to Lemma \ref{lem:stability}-\ref{e:ISS}, we get
		\begin{align*}
		V_m(\st_{t+1}) & \leq \gamma V_m(\st_t) +  c_3 \norm{g(\st_t)\tilde{\control}_t} \leq \gamma V_m(\st_t) + c_3 \frac{c_1}{c_2}\bar{c} \\
		& = (1-\frac{c_1}{c_2}) V_m(\st_t) + \frac{c_1}{c_2}\bar{c}.  
		\end{align*}
		Since $c \geq V_m(\st_t) > \bar{c}$ for all $\st_t \in \mathcal{X}_c(0)\setminus\mathcal{X}_{\bar{c}}(0)$, we have $V_m(\st_{t+1}) - V_m(\st_t) \leq \frac{c_1}{c_2} (\bar{c} - V_m(\st_t)) < 0$. 
		\item If $V_m(\st_t) \leq \bar{c}$ then $V_m(\st_{t+1}) \leq \gamma V_m(\st_t) +  c_3 \norm{g(\st_t)\tilde{\control}_t} \leq \gamma \bar{c} + \frac{c_1}{c_2}\bar{c} = \bar{c} \implies \st_{t+1} \in \mathcal{X}_{\bar{c}}(0)$.
		\item For every $\st_t \in \mathcal{X}_c(\st_t^r)\subset \mathcal{X}_c \subset \mathcal{X}$, $\st_{t+1} \in \mathcal{X}_c(\st_{t+1}^r) \subset \mathcal{X}_c \subset \mathcal{X}$ due to Proposition \ref{prop:main_tube}-\ref{prop:invariant_tube}.
	\end{enumerate}
\end{proof}
%%%%%%%%%%%%%%%%%
\begin{proof}[Proof of Theorem \ref{th:asymptotic}]
	Let us consider $V(\st_t, K_t) \Let V_m^2(\st_t) + a_0V_a(K_t)$, where $a_0 = \frac{2}{1-\theta} \left( \hat{c}_3 \hat{\delta}_g \sigma \right)^2$. Clearly, $V$ is continuous in $\st_t$ and $K_t$, and satisfies:
	\begin{equation*}
	\frac{a_0}{\theta} \trace(\tilde{K}_t\transp \tilde{K}_t) + c_1^2 \norm{\st_t}^4 \leq V(\st_t, K_t) \leq \frac{a_0}{\theta} \trace(\tilde{K}_t\transp \tilde{K}_t) + c_2^2 \norm{\st_t}^4.
	\end{equation*}
	for all $t \geq \hat{N} \geq N$. From Lemma \ref{lem:stability}-\ref{e:ISS} we have
	\begin{equation*}
	V_m^2(\st_{t+1}) \leq 2\gamma^2V_m^2(\st_{t}) + 2\hat{c}_3^2\norm{g(\st_t)\tilde{\control}_t}^2.
	\end{equation*} 
	Therefore,
	\begin{equation}
	V_m^2(\st_{t+1}) - V_m^2(\st_{t}) \leq -(1-2\gamma^2)V_m^2(\st_{t}) + 2\hat{c}_3^2 \hat{\delta}_g^2 \norm{\tilde{\control}_t}^2.
	\end{equation}
	Now, we compute $V(\st_{t+1}, K_{t+1}) - V(\st_t, K_t)$ and substitute $V_a(K_{t+1}) - V_a(K_t) \leq -\frac{1-\theta}{\sigma^2} \norm{\tilde{\control}_t}^2 + \frac{1}{m^2}\norm{\varepsilon_j(\st_t)}^2$ from Lemma \ref{lem:adaptation}-\ref{lem:drift_Va} to get
	\begin{align*}
	V(\st_{t+1}, K_{t+1}) - V(\st_t, K_t) & \leq -(1-2\gamma^2)V_m^2(\st_{t}) + \frac{a_0}{m^2} \norm{\varepsilon_j(\st_t)}^2 + 2\hat{c}_3^2 \hat{\delta}_g^2 \norm{\tilde{\control}_t}^2 - a_0 \left(\frac{1-\theta}{\sigma^2} \right) \norm{\tilde{\control}_t}^2 \\
	& = -(1-2\gamma^2)V_m^2(\st_{t}) + \frac{a_0}{m^2} \norm{\varepsilon_j(\st_t)}^2 \\ 
 &\leq  -(1-2\gamma^2)c_1^2 \norm{\st_t}^4 + a_0 \left(\frac{\beta}{m} \right)^2 \norm{\st_t}^4 \\
 &= -\eta \norm{\st_t}^4,
	\end{align*}
	where $\eta = (1-2\gamma^2)c_1^2 - \frac{2}{1-\theta } \left( \sigma \hat{c}_3 \hat{\delta}_g\frac{\beta}{m} \right)^2 > 0$ because $\beta < \frac{c_1 m}{\sqrt{2}\sigma \hat{c}_3 \hat{\delta}_g}\sqrt{(1-2\gamma^2)(1-\theta)}$.
	Therefore,
	\begin{align*}
	\norm{\st_t}^4 &\leq \frac{1}{\eta} \left( - V(\st_{t+1}, K_{t+1}) + V(\st_t, K_t) \right).
	\end{align*}
	By summing from $t=\hat{N}$ to $k+\hat{N}$ on both sides, we get
	\begin{align*}
	\sum_{t=\hat{N}}^{\hat{N} + k} \norm{\st_t}^4 & \leq  \frac{1}{\eta} \left( - V(\st_{\hat{N}+k+1}, K_{\hat{N}+k+1}) + V(\st_{\hat{N}}, K_{\hat{N}}) \right) \\
	& \leq \frac{1}{\eta} V(\st_{\hat{N}}, K_{\hat{N}}) = \frac{1}{\eta} \left( V_m^2(\st_{\hat{N}}) + a_0 V_a(K_{\hat{N}})\right)\\
	& \leq \frac{1}{\eta} \left( \hat{c}^2 +  \frac{4a_0}{\theta} \bar{W} \right), 
	\end{align*}
	where the last inequality is due to Lemma \ref{lem:adaptation}-\ref{lem:bound_Va} and the fact that $\st_{\hat{N}} \in \mathcal{X}_{\hat{c}}(0)$.
	Since the right hand side of the above inequality is independent of $k$, we have $\sum_{t=\hat{N}}^\infty \norm{\st_t}^4 \leq \frac{1}{\eta} \left( \hat{c}^2 +  \frac{4a_0}{\theta} \bar{W} \right)$, which implies $\norm{\st_t} \rightarrow 0$ as $t \rightarrow \infty$.
\end{proof}

\end{document}